\documentclass[3p,11pt,a4paper]{elsarticle}

\makeatletter
\def\ps@pprintTitle{}
\makeatother

\usepackage{xcolor}
\usepackage{lineno}
\usepackage{graphicx}
\usepackage[hidelinks]{hyperref}
\usepackage{amsmath}

\usepackage{amssymb}

\usepackage{pdfpages}
\begin{document}
\includepdf[pages=1,pagecommand={}]{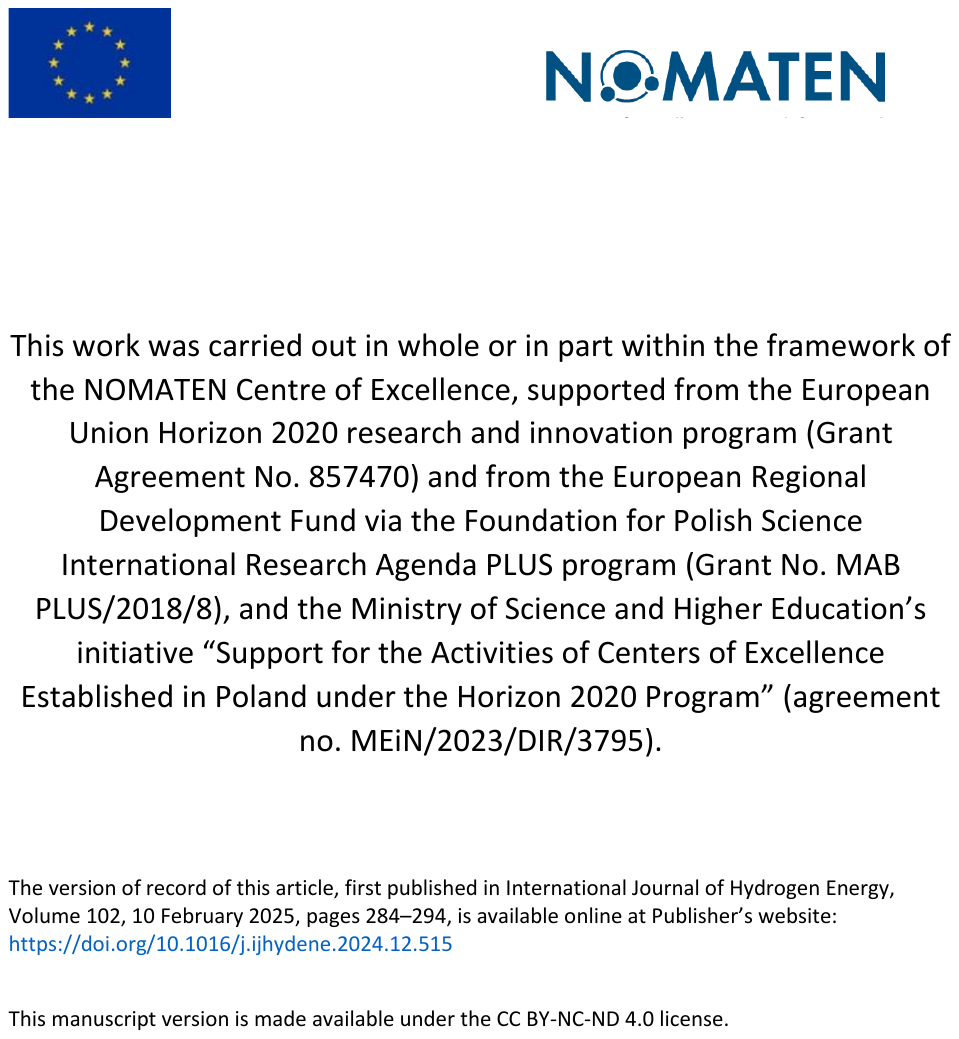}
\setcounter{page}{1}
\begin{frontmatter}




\title{Computational Study of Water Adsorption and Dissociative 
Mechanisms Impacting g--C$_3$N$_4$’s Optical and Electronic 
Properties}


\author[a,b,c]{Amil Aligayev}
\author[d]{Ulkar Jabbarli}
\author[c]{F. Javier Dom\'inguez-Guti\'errez}
\cortext[author] {Corresponding author: huangq@ipp.ac.cn}
\author[e,f]{Ulkar Samadova}
\author[a,b]{Jialin Li}
\author[c]{Stefanos Papanikolaou}
\author[a,b]{Qing Huang\corref{author}}
\address[a]{Key Laboratory of High Magnetic Field and Iron Beam Physical Biology, Institute of Intelligent Machines, Hefei Institutes of Physical Science, Chinese Academy of Sciences, Hefei 230031, China}
\address[b]{Science Island Branch of Graduate School, University of Science and Technology of China, Hefei, 230026, China}
\address[c]{NOMATEN Centre of Excellence, National Centre for Nuclear Research, ul. A. Sołtana 7, 05-400 Otwock, Poland}
\address[d]{Institute of Theoretical Physics, Faculty of Physics, University of Warsaw, Pasteura 5, PL-02093 Warsaw, Poland}
\address[f]{Institute of Physics Ministry of Science and Education Republic of Azerbaijan, H.Javid 131, AZ1143}
\address[e]{University of Electronic Science and Technology of China, Chengdu, 05-400, PR.China}

\begin{abstract}
In the quest for sustainable energy solutions, water splitting emerges as a crucial process for generating clean hydrogen—a versatile and renewable fuel essential for energy storage, emissions reduction, and achieving sustainability goals. This study employs a comprehensive computational approach, utilizing atomistic simulations to systematically investigate the effects of water absorption on the electronic and optical properties of g-C$_3$N$_4$ nanosheets.
Our methodology integrates ab initio computations grounded in density functional theory (DFT), which allows for a detailed characterization of the nanosheet and serves as a benchmark for self-consistent charge density functional tight binding (SCC--DFTB) simulations. This approach provides valuable insights into the behavior of the nanosheet under the influence of absorbed OH and H$_2$O molecules by considering calculated parameters for photocatalytic efficiency. Additionally, we extend our investigation to classical molecular dynamics simulations within the ReaxFF framework, modeling the emission of multiple H$_2$O molecules and assessing the subsequent rate of H$_2$ evolution.
A key finding of our study reveals that the dissociation of H$_2$O into HO and O molecules significantly enhances both the optical absorbance and conductance of the nanosheet compared to its pristine state. These results underscore the potential of g-C$_3$N$_4$ nanosheets as effective materials for water splitting applications.
\end{abstract}



\begin{keyword}
water splitting, DFT, 2D materials, g-C$_3$N$_4$, reaxFF, dftb, multiscale simulations.
\end{keyword}

\end{frontmatter}

\section{Introduction}
\label{sec:intro}

The interest in g-C$_3$N$_4$, a conjugated polymeric n-
semiconductor, has driven research into its photocatalytic and energy 
applications, spurring investigations of various modification 
strategies to improve its efficacy \cite{wang2024heterointerface, 
zhangband,ZHANG2024118884,LI2024114170,wang2024463,Algara}. The band-gap engineering and the development of 2D ultra-
thin nanosheets hold particular promise due to their simple preparation 
methods and controllable regulatory conditions \cite{ma2022band, 
wang2022high}. Furthermore, DFT is frequently used to study the
properties of materials and to elucidate experimental phenomena
\cite{humayun2023enhanced, mehtab2023photo}. 
Understanding the interaction between g-C$_3$N$_4$ and water (H$_2$ O) is crucial to elucidate the water splitting mechanism in photocatalysis \cite{wu2014effect, ma2011adsorption, liu2011initial,SHI2024114018,MORTAZAVI202040}. However, there is limited information on this interaction. More research is needed to clarify how H$_2$O affects the electronic properties of g-C$_3$N$_4$, including its electron transport probability, optical absorption properties, band structure and band edge position, which are essential for optimizing its performance as a photo--catalyst.

Photocatalysis technology has garnered significant attention for its extensive applications in environmental remediation, energy production, and photochemical synthesis, primarily due to its environmentally friendly, clean, and zero-pollution nature \cite{SONG2024135132}. A key advantage of photocatalytic water splitting is its capacity to directly convert abundant and renewable solar energy into chemical energy stored as hydrogen fuel. Hydrogen, being a clean and carbon-free energy carrier, can be utilized in diverse applications, such as fuel cells for electricity generation, transportation, and industrial processes, without emitting greenhouse gases \cite{zhong2022defect, lian2022harnessing,li2023localized, gao2024modulation}. 
In the realm of photocatalytic hydrogen evolution, various 
semiconductors such as, TiO$_2$ \cite{alsalka2018co, wang2024dual}, CdS \cite{ma2023broadened, xu2022phosphorus}, 
ZnIn$_2$S$_4$ \cite{xi2024electron, sun2024interfacial}, graphitic 
carbon nitride (g--C$_3$N$_4$) 
\cite{fu2018g,cong2024role,WEN201772,Algara} and others have been 
extensively studied. However, due to non-toxicity, ease of 
preparation, low--cost, chemically stable and suitable bandgap g-C$_3$N$_4$ stands out as a promising photocatalyst in all kinds of carbon nitride \cite{fu2018g, yan2024immobilization, mahvelati2023coordinatively}. Carbon nitride comprises six distinct phases: $\alpha$--C$_3$N$_4$, $\beta$--C$_3$N$_4$, cubic-C$_3$N$_4$, pseudocubic--C$_3$N$_4$, amorphous --C$_3$N$_4$, and g--C$_3$N$_4$. Among these phases, g--C$_3$N$_4$ is the most stable, characterized by a graphene-like layered structure. 

The prediction by Liu and Cohen that carbon nitride could serve as a unique material has 
catalyzed significant research on g-C$_3$N$_4$ \cite{liu1989prediction}. Gillan et al. has 
highlighted its exceptional stability and chemical resistance \cite{gillan2000synthesis}. X. Zhang et al have noticed its applications to facilitating solar energy harvesting \cite{D4EN00399C} and the importance of noble metals that are relevant nanocomponents for g-C$_3$N$_4$ used in water splitting \cite{Zhang2024}; as well as the photocatalytic evolution of H$_2$ and H$_2$O$_2$ for clean energy application \cite{TONG2023116805}. Zhou et al. comprehensively reviewed its synthesis and applications in 2019 
\cite{zhou2017preparation}. Through DFT computational studies, 
Kroke et al. elucidated that the tri-s-triazine-based configuration of g-C$_3$N$_4$ possesses an 
energy 30 kJ mol$^{-1}$ lower than its s-triazine-based counterpart, indicating superior 
thermodynamic stability \cite{kroke2002tri}.
Consequently, these findings have precipitated a significant surge in research initiatives focused on tri-s-triazine g-C$_3$N$_4$, highlighting its 
increasing prominence in scientific discourse.

Understanding water molecule interactions with materials like 
graphitic carbon nitride (g-C\(_3\)N\(_4\)) requires a 
multiscale modeling approach to capture phenomena occurring
at diverse time and length scales. In this study, we 
integrate density functional theory (DFT), 
self--consistent charge density functional tight binding
(SCC--DFTB), and reactive force field (ReaxFF) molecular 
dynamics, complemented by van der Waals corrections.
Each method brings unique strengths and limitations.
DFT provides atomistic accuracy and insight into
electronic structure but is computationally intensive,
limiting its application to small systems and short time 
scales. SCC-DFTB offers a balance between accuracy and 
efficiency, extending simulations to larger systems
while capturing essential electronic interactions. 
However, it sacrifices some precision in chemical bonding 
details. ReaxFF enables molecular dynamics over larger
length and time scales to study dynamic processes but
lacks the quantum mechanical rigor needed for electronic 
property predictions. This methodological combination
ensures that limitations of individual approaches are 
mitigated, offering a holistic perspective on adsorption, 
dissociation, and molecular formation dynamics.
In the present study, we applied a multi--scale modelling 
based on DFT, SCC-DFTB and ReaxFF calculations to 
elucidate the interactions between H$_2$O molecules and
the tri-s-triazine-based g-C$_3$N$_4$ 
surface, as well as the dissociative mechanisms of 
water molecules.
The findings of this research provide a robust theoretical framework 
for the application of g-C$_3$N$_4$ as a promising photocatalyst, potentially advancing the field of 
sustainable energy production.

\section{Computational methods}
\label{sec:methods}

Density Functional Theory (DFT) is a quantum mechanical method
used to study the electronic structure of atoms, molecules,
and solids by focusing on electron density rather than
wavefunctions. Based on the Hohenberg-Kohn theorems, DFT
asserts that all ground-state properties of a system are
determined by its electron density, which minimizes the
system's total energy. 
Thus, we conduct DFT computations using the Vienna Ab initio
Simulation Package (VASP) \cite{PhysRevB.47.558} to accurately
describe the electron--ion interactions with the projector-augmented 
wave (PAW) method \cite{kresse1996efficient}, while considering the 
valence electrons of C 2s$^2$2p$^2$, N 2s$^2$2p$^3$ and O 2s$^2$2p$^4$. 
For the exchange-correlation functional, we applied the generalized 
gradient approximation (GGA) \cite{perdew1996generalized} in 
the Perdew-Burke-Ernzerhof (PBE) formulation \cite{perdew1992atoms}.
The electronic wavefunctions were expanded in a plane wave basis set 
with a cutoff energy of 460 eV. The weak van der Waals interactions
were accounted for using the DFT-D3 approach with the 
Becke--Johnson scheme \cite{grimme2010consistent}. 
Thus, we can benchmark calculations for the interaction of H\(_2\)O 
molecules with the 2D sheet, which is essential for validating
larger-scale simulations using approximate methods to DFT.

\subsection{SCC--DFTB method}
\label{subsec:dftb}

The SCC-DFTB method is a computational approach that is rooted
in the theoretical foundation of DFT, specifically in
the Kohn-Sham (KS) framework, where the energy is articulated
through a linear combination of atomic orbitals (LCAO) over
a minimal basis set. A crucial aspect involves approximating
this quantity through a Taylor expansion concerning
a reference density, truncated at various orders, thus
establishing a hierarchy of DFTB methods \cite{elstner1998self,gausdftb}.
Utilizing Slater--Koster (SK) parameter files, 
it offers tabulated Hamiltonian matrix elements, overlap 
integrals, and repulsive splines fitted to DFT dissociation 
curves. Notably, the mio SK parameters are employed due to their optimization for modeling light elements.
They offer accurate descriptions of bonding 
interactions and energy landscapes while providing a 
reliable representation of the potential energy 
surface.
These parameters detail the overlap and hopping integrals
between pairs of atoms in the tight--binding Hamiltonian.
Achieving an optimal set of SK parameters
requires two primary criteria: accurately reproducing the 
structure of relevant electronic bands and faithfully
representing the orbital contribution within those bands 
\cite{DFTBplus, Qiang}.
The SCC-DFTB a second-order manifestation of the KS
energy, has demonstrated notable success in describing
the physical and chemical processes involved in the interactions
of materials and molecules within Molecular Dynamics
(MD) simulations \cite{gausdftb,aligayev2024computational,C7TC00976C,DOMINGUEZGUTIERREZ2018189}. 
Its precision, when juxtaposed with
experimental values, mirrors that of comprehensive DFT
calculations conducted with a double-${\zeta}$ plus
polarization basis set. It can precisely predict the
structures and thermodynamic properties, providing insights
into the gas adsorption interaction with 
2D carbon-based materials and their potential applications
in processes such as water splitting and dissociative mechanisms.


Therefore, in the scope of this approach the total energy
of the system is expressed as:
\begin{equation}
    E^{\rm DFTB} = E_{\rm band}+E_{\rm rep}+E_{\rm SCC},
\end{equation}
where the band structure energy,
\(E_{\text{band}}\), is derived from the summation of orbital
energies ($\epsilon_i$) 
over all occupied orbitals ($\Psi_i$). 
The repulsive energy, ($E_{\text{rep}}$), concerns 
core--core interactions, including exchange--correlation
energy and other contributions formulated in 
distance--dependent pairwise terms. 
The SCC contribution, ($E_{\text{SCC}}$), encompasses 
charge-charge interactions in a system. 
Herein, we employed the mio SK parameter set, a well-established 
parameterization scheme for SCC-DFTB calculations that has been
used in our previous work to describe the interactions
among C, N, H, and O atoms \cite{DOMINGUEZGUTIERREZ2018189, aligayev2024computational, gausdftb}. 
This parameter set, implemented with dispersion corrections,
ensures accurate representation of the electronic structure
and interatomic interactions in the given
system \cite{elstner1998self}. 
The SCC-DFTB approach is used here to model the interaction of water molecules with the 2D material and to investigate its effects on the optical and electronic properties of g-C\(_3\)N\(_4\). This method enables the study of larger systems than those typically employed in DFT calculations.

\subsection{ReaxFF approach}
\label{subsec:reaxff}

While the SCC--DFTB method proves to be a powerful tool for the 
exploration, development, and optimization of novel materials,
its computational cost in semi--classical molecular dynamics
simulations at a larger scale for periodic systems is a limiting factor.
This limitation hampers the consideration of dynamic
system evolution, preventing simulations that closely mimic
experimental conditions for large scale atomic system, e.g. hundred or thousands of atoms \cite{LOPEZPLASCENCIA20174774,HUANG2019107850}.
To overcome this challenge, empirical force fields, trained
using DFT structure and energy data, are employed to facilitate 
simulations with significantly reduced computational demands. One prominent method for this purpose is the reactive force field
(ReaxFF), as implemented in the Large-scale Atomic Molecular Massively Parallel 
Simulator (LAMMPS) \cite{LAMMPS}; while sacrificing some accuracy, provides
a substantial decrease in computational expense 
\cite{senftle2016reaxff,vanDuin,Rappe}.
In our work, we leverage reactive force fields like ReaxFF,
incorporating connection--dependent terms that describe reactive
events through a bond--order formalism derived from interatomic
distances to investigate dissociative mechanisms of molecules colliding with the 2D material. This characteristic enables ReaxFF to simulate reaction 
chemistry without explicit DFT consideration, allowing for the 
exploration of phenomena previously inaccessible through computational 
methods \cite{senftle2016reaxff}. ReaxFF utilizes a bond-order formalism 
alongside polarizable charge descriptions to model both reactive and non-
reactive interactions between atoms accurately. This capability extends 
to covalent and electrostatic interactions for g-C$_3$N$_4$ and H$_2$O 
molecules.

The energy contributions to the ReaxFF potential are summarized by the 
following terms: 
\begin{align}
    E_T & = E_{\rm bond} + E_{\rm angle} + E_{\rm tors} \nonumber \\
        & + E_{\rm over} + E_{\rm Coulomb} + E_{\rm vdWaals}
        + E_{\rm specific}.
\end{align}
Here, $E_{\rm bond}$ is a continuous function of interatomic distance, 
representing the energy associated with forming bonds between atoms. 
$E_{\rm angle}$ and $E_{\rm tors}$ account for three--body valence angle 
strain and four-body torsional angle strain, respectively. $E_{\rm over}$ 
imposes an energy penalty to prevent the over-coordination of atoms, 
based on atomic valence rules. $E_{\rm Coulomb}$ and $E_{\rm vdWaals}$ 
contribute electrostatic and dispersive interactions between all atoms, 
irrespective of connectivity and bond-order. Finally, $E_{\rm specific}$ 
encompasses system-specific terms, selectively included to capture unique 
properties of the system, such as lone-pair effects, conjugation, and 
hydrogen bonding.

\subsection{Characterization of g-C$_3$N$_4$}

\begin{figure}[t!]
    \centering
    \includegraphics[width=0.48\textwidth]{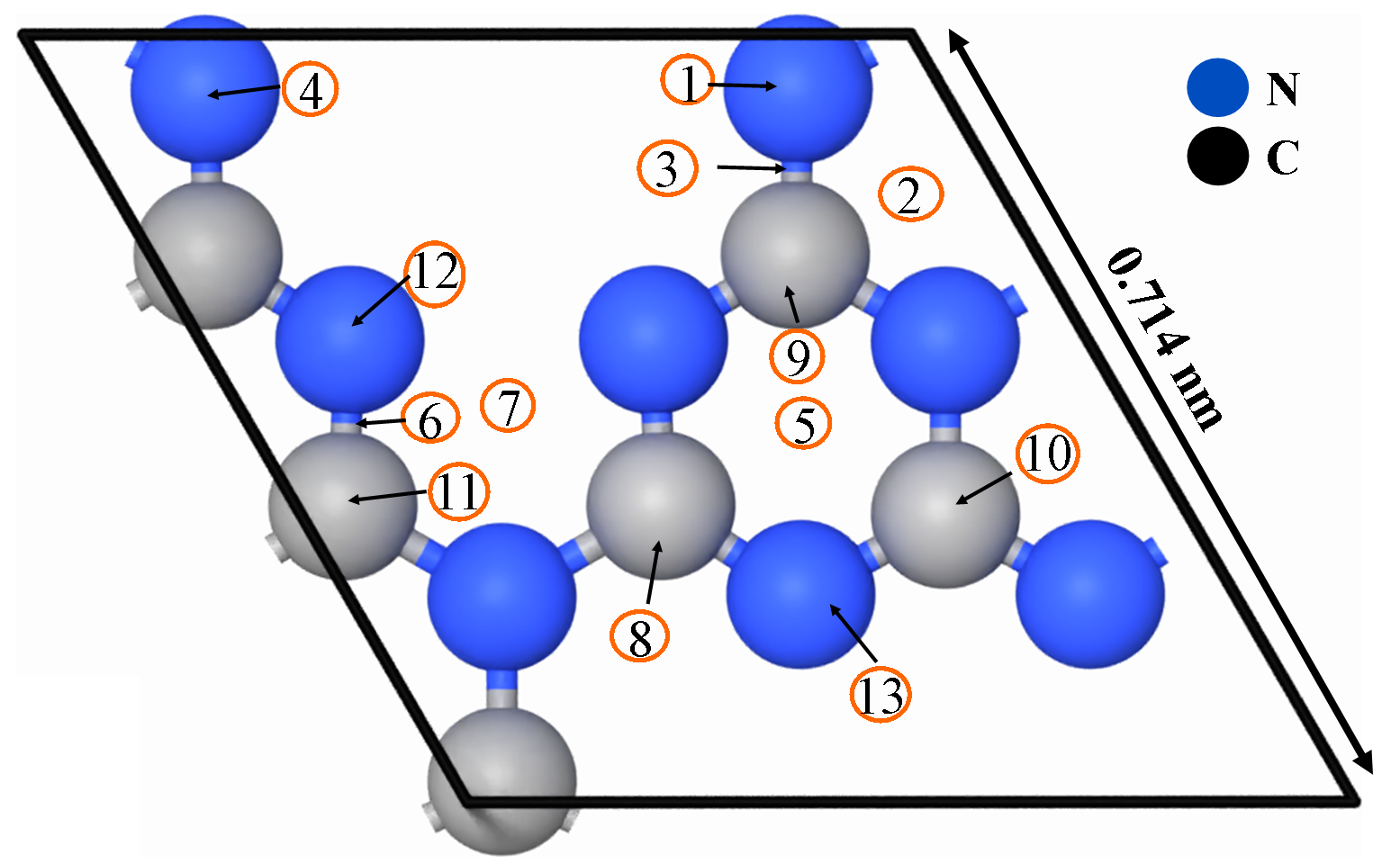}
    \caption{Optimized structures of g--C$_3$N$_4$ using DFT,  
    SCC--DFTB, and reaxFF approaches. 
    The calculated bond lengths and lattice parameters 
    are in good agreement with the reported DFT results \cite{HEnda}. 
    Adsorbate sites considered for potential energy curves are 
    enumerated for further physisorption pathways calculations.}
    \label{fig:AdsorbPEcs1}
\end{figure}

The pristine g-C$_3$N$_4$ sheet is derived from a 
unit cell oriented along the (001) direction, as illustrated
in Fig. \ref{fig:AdsorbPEcs1}. The optimized structure
consists of six carbon atoms and eight nitrogen atoms. 
In the DFT computations, the total energy difference criterion
in the self--consistent field loop was set to 10$^{-6}$ eV,
while geometry optimizations were terminated when the
force on each atom fell below 0.01 eV/\AA. 
In addition, a Monkhorst--Pack grid \cite{monkhorst1976special}
of $15\times15\times1$ is utilized for Brillouin zone sampling. For the SCC-DFTB calculations, the geometry optimization followed 
the same methodology for DFT one, while the 2D materials were
energy relaxed with ReaxFF by using FIRE minimization algorithm
until the force reach $1 \times 10^{-6}$ eV/\AA{} \cite{GUENOLE2020109584} allowing the samples to find their 
lowest energy structure.
The lattice parameters, with $a = b = 0.715$ \AA{}, 
and internuclear distances obtained through DFT, SCC--DFTB, 
and reaxFF approaches are found to be in accordance with
consistent findings from other computational results \cite{HEnda,D0RA09815A}.
By this unit cell, we identify several adsorbate sites 
for further potential energy curves calculations.

\begin{figure}[t!]
    \centering
    \includegraphics[width=0.48\textwidth]{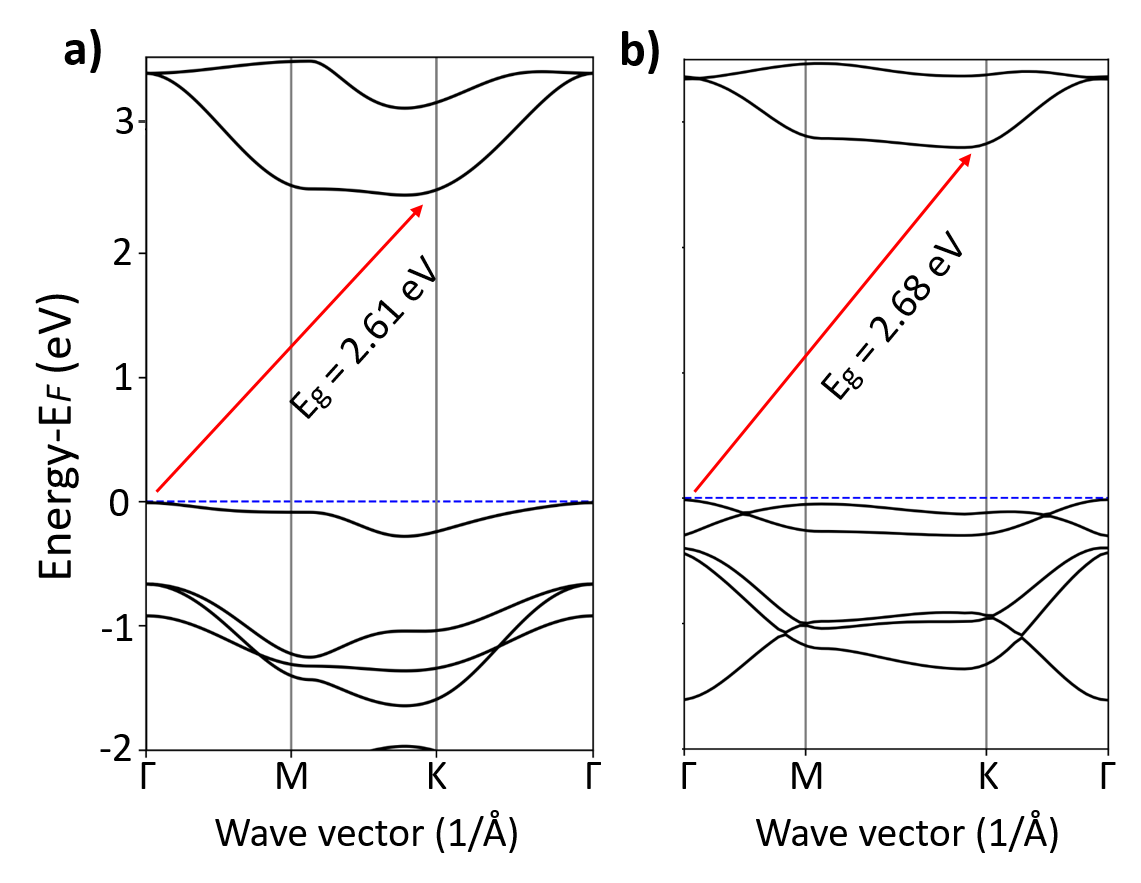}
    \caption{Band structure calculated using DFT and DFTB approaches, showing good agreement in the band gap values of 2.61 eV and 2.68 eV, respectively. Despite inherent differences in the energy bands at negative energies, which are attributed to the approximation of the Slater-Koster (SK) parameters, both methods provide consistent results for the band gap.}
    \label{fig:band-structure}
\end{figure}

In addition, we calculate the band structure of the g-C$_3$N$_4$ monolayer using both DFT and DFTB 
approaches, as illustrated in Fig. \ref{fig:band-structure}, revealing its indirect band gap nature.
According to DFT calculations, the band gap (E$_g$) is 
found to be 2.61 eV, while the DFTB method yields a 
slightly higher value of 2.68 eV, which is in good 
agreement with several experimental studies \cite{MOLAEI202332708,ong2016graphitic,mishra2019graphitic, akple2015enhanced} and numerical calculations\cite{D0RA09815A,XU201211072}. The hybridization of carbon and nitrogen atoms 
contributes to the formation of well-defined 
conduction and valence bands. In both cases, the conduction band minimum (CBM) is located at the K point, while the valence band maximum (VBM) is positioned near the $\Gamma$ and M points, respectively. This discrepancy between the DFT and DFTB results arises due to the differing levels of approximation in each method.

\section{Physisorption pathways}
\label{sec:results}

The interaction potentials between H\(_2\) and H\(_2\)O
molecules and the g-C\(_3\)N\(_4\) sheet were investigated
using adiabatic calculations to determine potential
energy curves. For the DFT and SCC--DFTB calculations,
a self--consistent approach was employed, while a single 
energy computation was performed using ReaxFF for the
molecules and the fully relaxed sheet within a 2\(\times\)2
supercell at various distances and adsorption sites.
Van der Waals interactions were accounted for by
including dispersion corrections in all calculations 
\cite{10.1063/1.1329889}; 
a vacuum space of at least 15 \AA{} is employed to 
prevent interactions between periodic images.
Then, the total energy, denoted as $E(z)$, pertain 
to the system of molecule--2D material at separations $z$ 
ranging from 0.5 to 7 \AA{} above the surface. 
This span helps
define the computation of the adsorption potential
concerning the distance separation.
The total energy is then computed as:
\begin{equation}
    E(z) = E_{\rm Tot} - \left( E_{\rm 2D} + E_{\rm Molecule}
    \right),
\end{equation}
The notations used in the calculations are:
$E_{\rm 2D}$, representing the total energy of the
g-C$_3$N$_4$; $E_{\rm Molecule}$, signifying the total
energy of the isolated molecule types: H$_2$ and H$_2$O; 
and $E_{\rm Tot}$, referring to the system's energy at
each $z$-distance. 
The binding energy, denoted as
$E_b$, is derived from $E(z_{\rm min})$, with $z_{\rm min}$ 
denoting the equilibrium distance between the molecule
and the surface. The computation involves evaluating 
the total energy of the system, specifically assessing 
13 unique adsorption sites for g--C$_3$N$_4$ (see Fig. 
\ref{fig:AdsorbPEcs1} for visualization on the 
optimized unit cell of the g--C$_3$N$_4$ sheet) while 
accounting for the unit cell configuration. This 
process includes adjusting the distance between the 
surface and the center of mass of the molecules along 
the $z$-axis.

\begin{figure}[b!]
    \centering
    \includegraphics[width=0.48\textwidth]{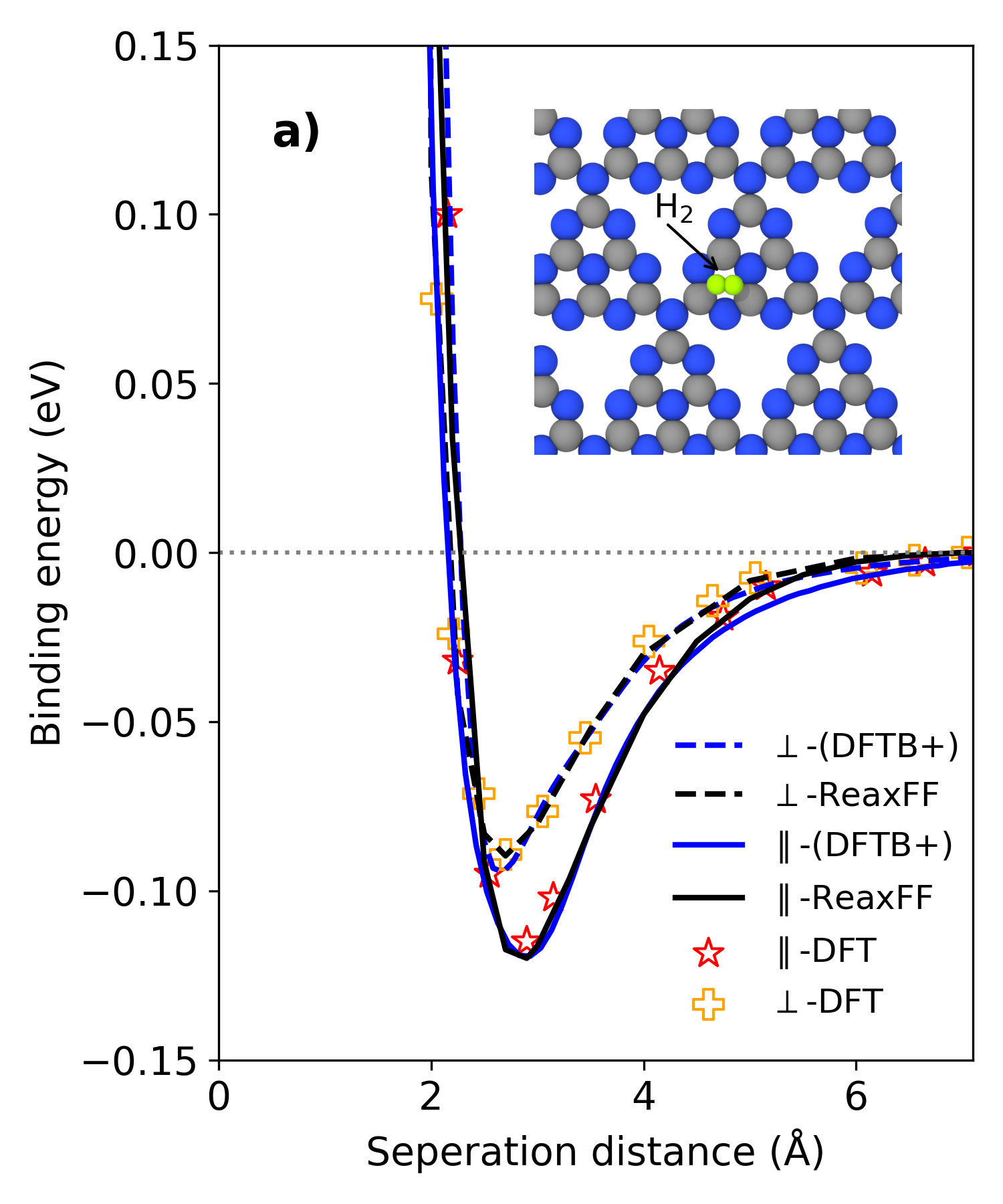}
    \caption{Physisorption pathways of the hydrogen  molecule on
    g-C$_3$N$_4$, showing the minimum binding energy
    from 13 adsorbate sites at the hole of the hexagonal ring, 
    as shown in the inset figure.
    The orientations for the H$_2$ molecules include both 
    parallel ($\parallel$) and perpendicular ($\perp$) 
    configurations relative to the sheet. We notice a good 
    agreement across methods.}
    \label{fig:PECsH2}
\end{figure}

In Fig. \ref{fig:PECsH2}, we present the physisorption
pathways for the H$_2$ molecule at the minimum binding 
energy selected from all of the 13 potential sites,
considering both parallel ($\parallel$) and perpendicular
($\perp$) orientations of the H--H bond relative to the
g--C$_3$N$_4$ sheet. The results from DFT (VASP), 
SCC-DFTB (DFTB+), and ReaxFF (LAMMPS) calculations show
strong agreement across methods. In the inset figure, 
we depicted the H$_2$ molecule reaching
a bond length of 3.1 Å at adsorption site 5 in the parallel
orientation, located at the center of the hexagonal ring,
likely due to $2p$ bonding interactions between C and N atoms.
The perpendicular orientation yields results similar to those
of a single hydrogen atom interacting with the surface.
This consistency across different computational approaches
enables reliable dynamic modeling of hydrogen molecule
formation from water molecule interactions with the 2D 
sheet, as well as accurate predictions of the vibrational
states of hydrogen molecules.

\begin{figure}[t!]
    \centering
    \includegraphics[width=0.48\textwidth]{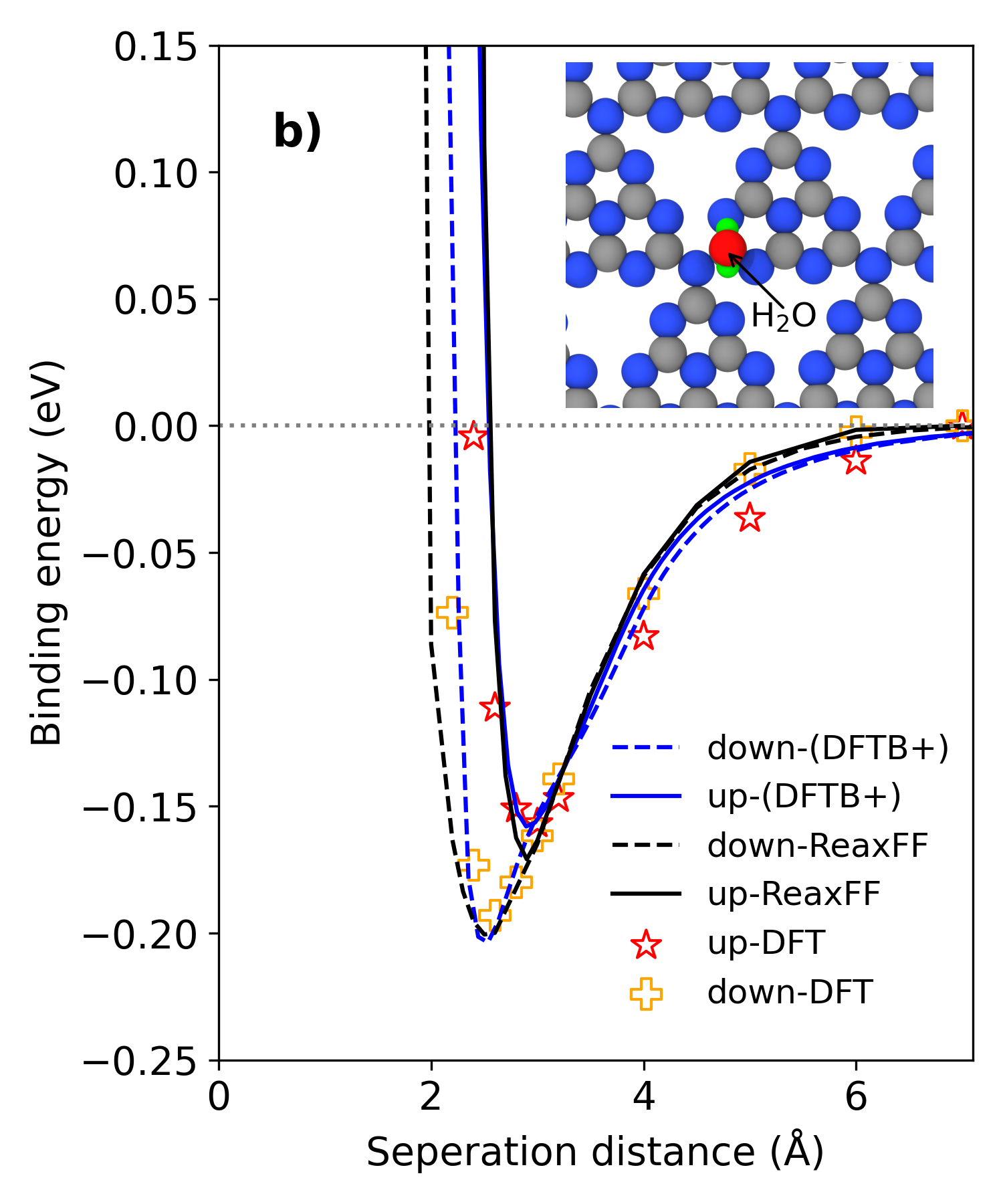}
    \caption{Physisorption pathways of the water  molecule on
    g-C$_3$N$_4$, showing the minimum binding energy
    from all adsorbate sites above the C atom at the left lower 
    corner of the triangle of the sheet for the the \(\sigma(xz)\) 
    plane of the molecule (labeled as donw), as shown in the inset figure. Good 
    agreement is 
    obtained among the approaches.}
    \label{fig:PECsH2O}
\end{figure}

In Fig. \ref{fig:PECsH2O}, we present the physisorption pathways
for the H\(_2\)O molecule. Given that water and its four symmetry 
elements form the point group C\(_{2v}\), we consider two symmetry
configurations: "down," where the two hydrogen atoms are 
oriented toward the g-C\(_3\)N\(_4\) surface (corresponding
to the \(\sigma(xz)\) plane), and "up," where the oxygen atom
is closer to the sheet (corresponding to the \(\sigma(xy)\) plane).
In both configurations, the adsorption site labeled as 8 shows
the lowest energy. The hydrogen atoms are more strongly
attracted to the sheet due to the interplay between the $2p$
orbitals of the oxygen atom and the C and N atoms of the surface.
This configuration results in a separation distance
of 2.4 Å for the minium binding energy, as depicted in the inset figure.

\begin{figure}[t!]
    \centering
    \includegraphics[width=0.3\textwidth]{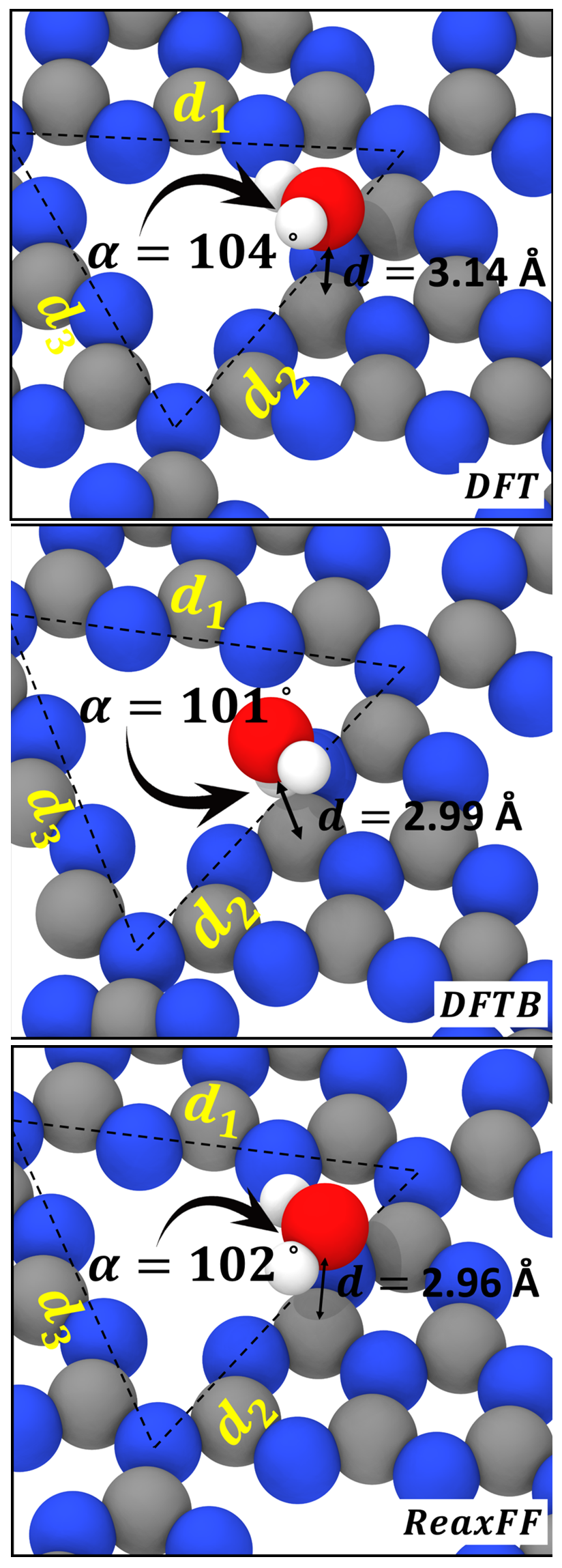}
    \caption{Optimized structures of g--C$_3$N$_4$ using DFT,  
    SCC--DFTB, and reaxFF approaches. 
    The calculated bond lengths and lattice parameters 
    are in good agreement with the reported DFT results \cite{HEnda}. }
    \label{fig:AdsorbPEcs}
\end{figure}

Once the binding energy between the molecules and the g-C$_3$N$_4$ 
sheets was determined, the next step involved optimizing the system's 
energy by positioning the H$_2$O and H$_2$ molecules at the identified 
adsorption sites. The optimization was carried out using DFT, SCC-DFTB approaches with the conjugate gradient method, while FIRE protocol was used for ReaxFF calculations. In all cases, 
the structures were fully relaxed with respect to volume, shape, and 
internal atomic positions until the atomic forces were reduced to less 
than $10^{-4}$ eV/Å for the entire numerical cell.
In Fig. \ref{fig:AdsorbPEcs}, we present the optimized structures
for the water molecule, noting that the H$_2$O molecule undergoes 
rotation during the optimization process for all approaches. The DFT 
and DFTB yield a final distance of 3.14 Å and 2.99 Å between the oxygen atom and 
the g-C$_3$N$_4$ sheet, respectively, with a configuration where one hydrogen atom is 
close to a C-atom in case of DFTB. In contrast, the reaxFF approach predicts a slightly 
shorter O-C distance of 2.96 Å, with the three atoms (O and two H) 
lying in a plane parallel to the 2D material.
This discrepancy highlights the limitations of the reaxFF framework, 
where charges are tabulated, in contrast to the DFT and DFTB 
approaches, which account for more comprehensive electronic 
contributions in the calculations.

\subsection{Spin Effects}
\label{subsec:spinEffects}

\begin{figure}[t!]
    \centering
    \includegraphics[width=0.48\textwidth]{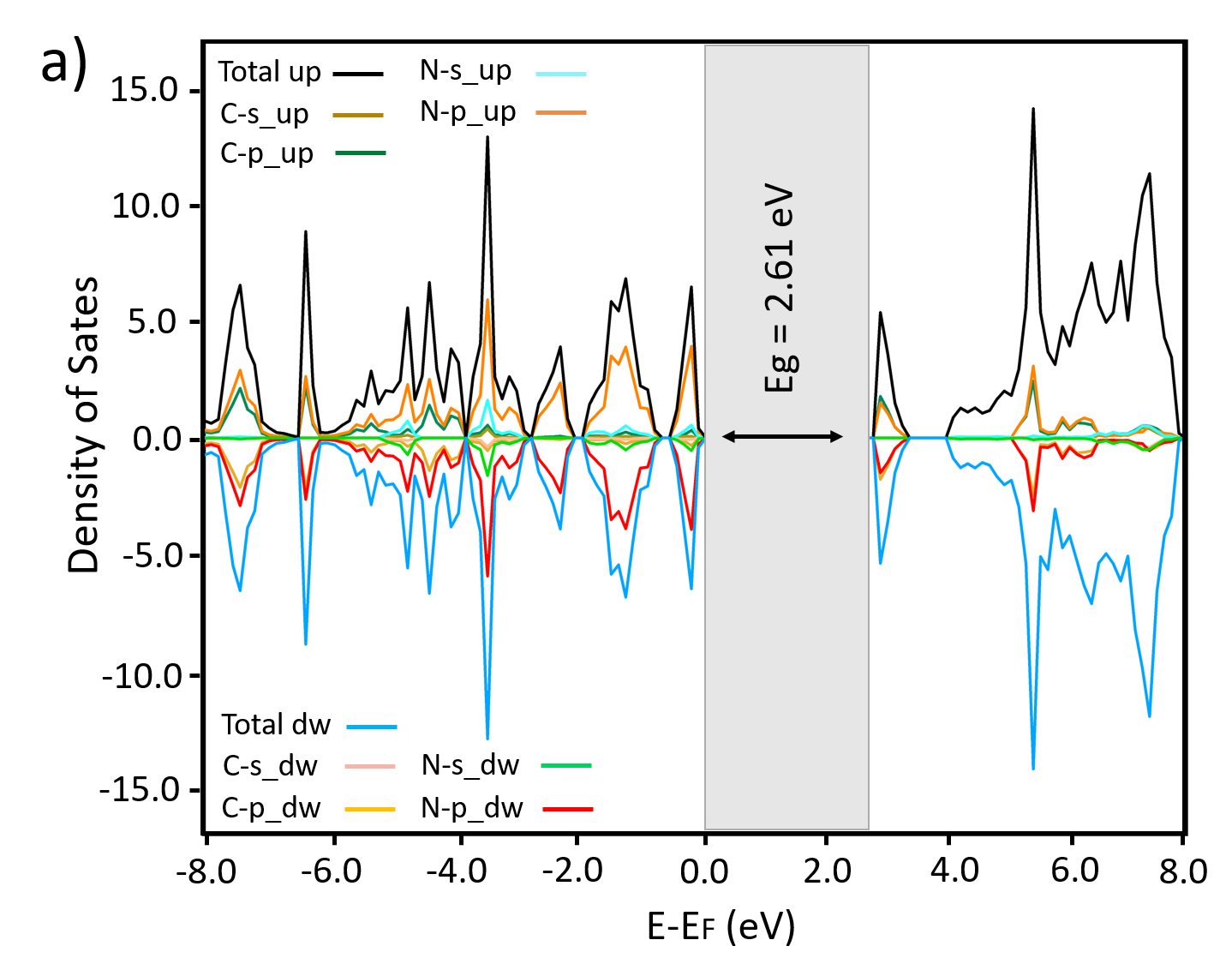}
    \includegraphics[width=0.48\textwidth]{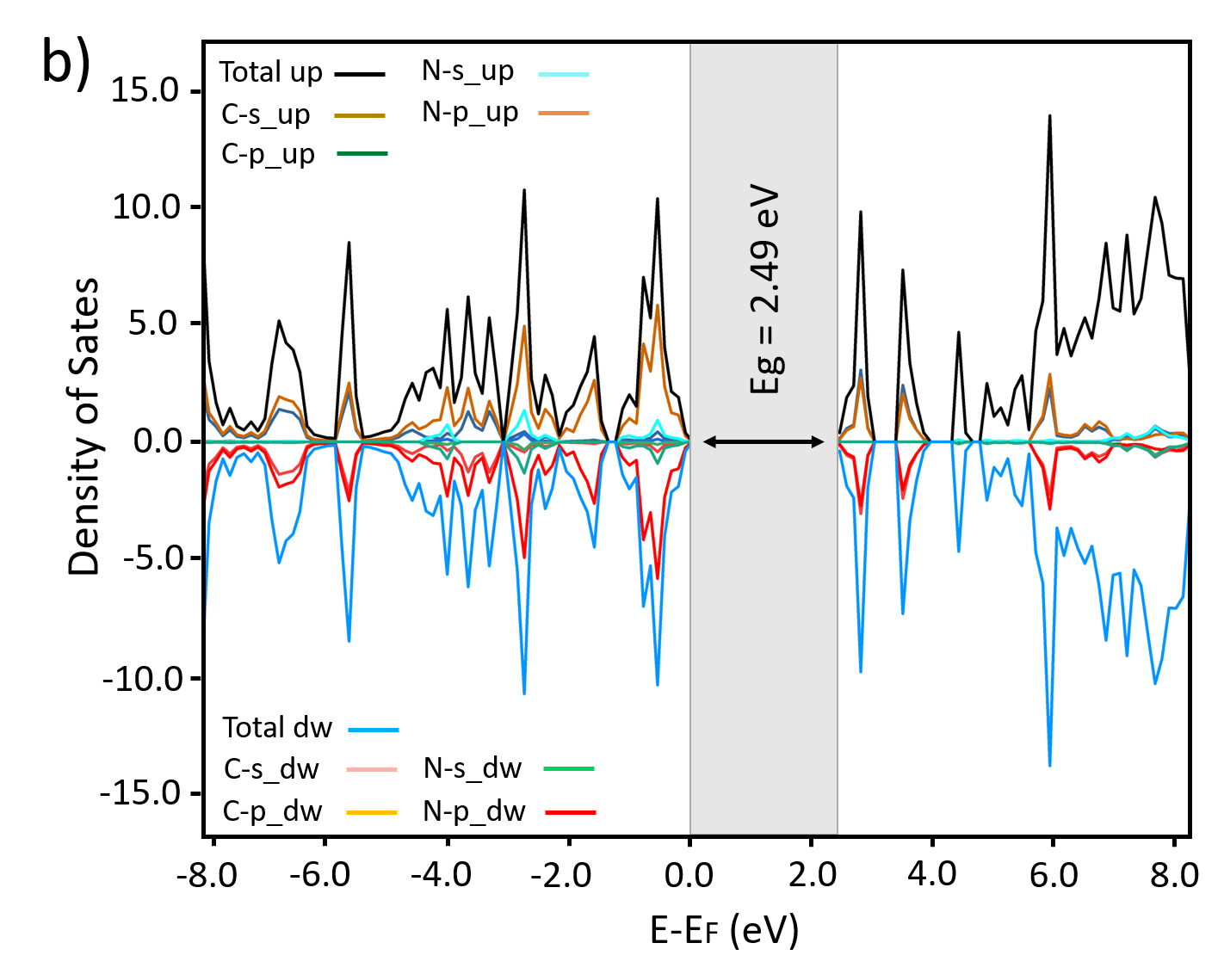}
    \caption{
    Density of States (DOS) for pristine g-C$_3$N$_4$ (a) and g-C$_3$N$_4$ with an attached H$_2$O molecule (b), considering spin effects. The pristine system exhibits no significant spin polarization, whereas the adsorption of the H$_2$O molecule induces localized spin effects, highlighting the influence of water attachment on the electronic structure. .}
    \label{fig:dos-spin}
\end{figure}

To gain a deeper understanding of the electronic properties of the studied 
material, we conducted spin--polarized calculations to compute the DOS. 
In Fig \ref{fig:dos-spin}a), we present DOS for the pristine g-C$_3$N$_4$, which shows no 
significant spin effects, consistent with its non-magnetic nature as reported \cite{gao2018halfmetallicity}. 
However, in Fig. \ref{fig:dos-spin}b) we shows the DOS 
upon the 
adsorption of H$_2$O molecules onto the g-C$_3$N$_4$ surface, distinct spin 
polarization becomes evident in the DOS. This suggests that the interaction with 
H$_2$O induces localized spin effects, likely due to charge redistribution and 
the breaking of spin symmetry in the system.
In addition, the band gap is decreased to 2.49 eV, respect to the pristine case of 2.61 eV.
This observation highlights the 
sensitivity of g-C$_3$N$_4$'s electronic properties to surface interactions and 
underscores the importance of spin considerations when modeling its behavior in 
functionalized or chemically modified environments.

\section{Optical and Electronic properties}
\label{sec:optica}

Optical absorption is explored within 
DFTB framework, treating it as an electronic dynamic process in
response to an external electric field \cite{C8CP04625E,B926051J}. 

The conventional adiabatic approximation is employed to obtain the
time evolution of the electron density matrix, achieved through the
time integration of the Liouville--von Newmann equation.
\begin{equation}
    i \hbar \frac{\partial \hat \rho}{\partial t} = 
    S^{-1}\hat H \hat \rho - \hat \rho S^{-1},
\end{equation}
where $\hat \rho$ is the single electron density matrix, 
$\hat S$ is the overlap matrix, and $\hat H$ is the system Hamiltonian 
that includes the external electric field as 
$\hat H = \hat H_0 + E_0 \delta (t-t_0) \hat e$ with 
$E_0$, the magnitude of the electric field, 
and $\hat e$, its direction. 
Within the context of linear response, the absorbance $I(\omega)$ is 
determined as the imaginary part of the Fourier transform of the induced
dipole moment induced by an external field. In our investigation, 
the external field strength was fixed at ($E_0 = 0.1$) V/\AA{}. The 
induced dipole moment was computed over a time span of (200) fs, employing 
a time step of ($\Delta t = 0.01$) fs. 
To mitigate noise, the Fourier transform was executed with an exponential
damping function, utilizing a damping constant of (5) fs.
The applied laser field is defined as:
\begin{equation}
    E(t) = E_0 f(t) \sin(\omega t+\phi),
\end{equation}
where $E_0$ is the strength of the electric field, $\omega$ is the 
laser frequency and $\phi$ the phase. We consider a Gaussian 
envelope function $f (t) = \exp(-(t-t_m)2/\beta^2)$, 
where $t_m$ is the time at which the pulse is 
centered and $\beta = \tau/2 \sqrt{\pi}$, $\tau$ being the duration
of the pulse. This allows us to model a laser pulse of 30 fs with 
an energy of 2.55 eV (486 nm).

\begin{figure}[t!]
    \centering
    \includegraphics[width=0.48\textwidth]{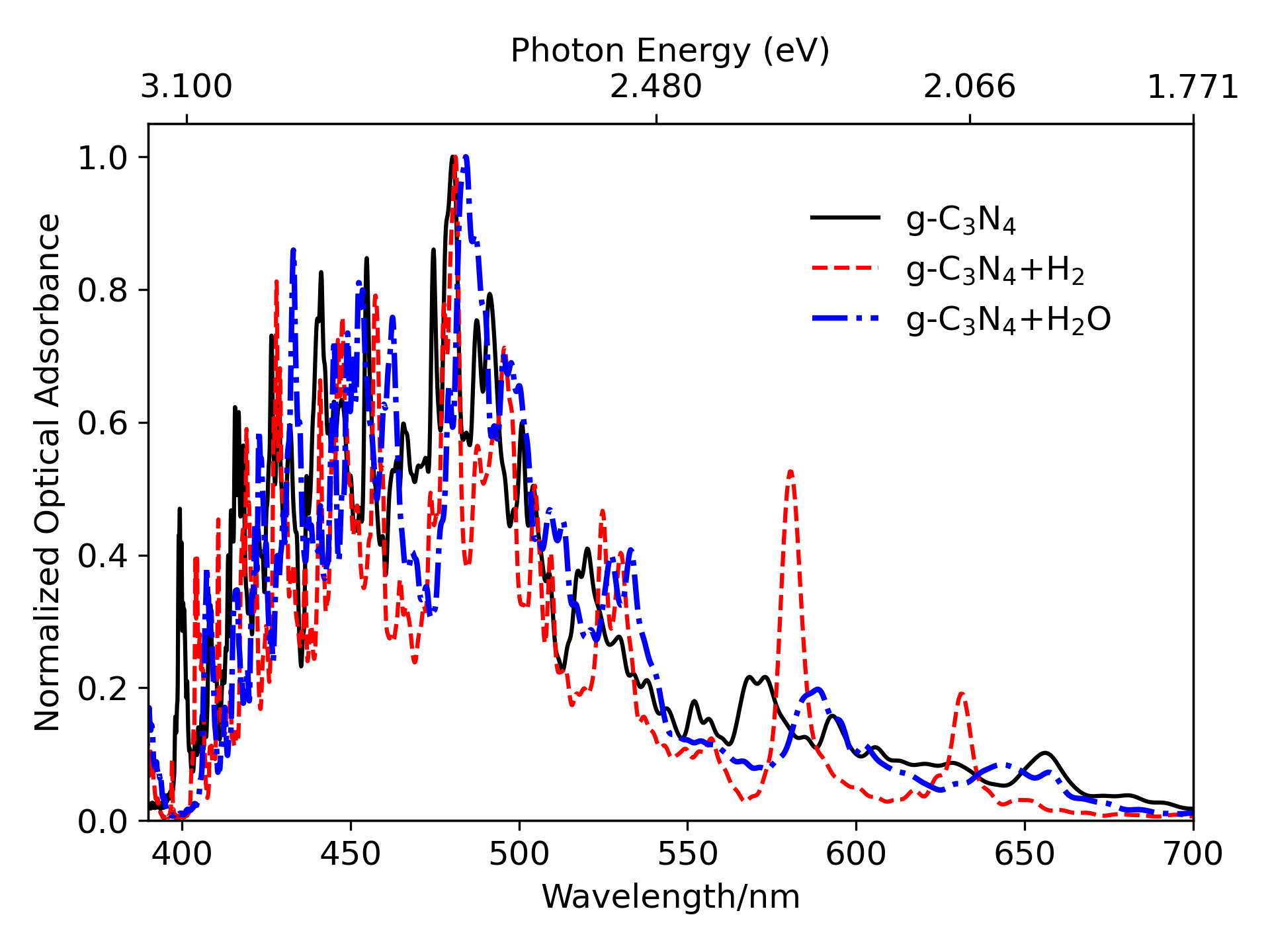}
    \caption{(Color online) Optical absorbance of g-C$_3$N$_4$ obtained experimental (thick gray line) and computationally (black line) in good agreement for the wavelenght range of 400 to 700 nm. Results for the H$_2$ and water molecule at the adsorbates sites are presented.}
    \label{fig:enter-label-opt}
\end{figure}

Fig. \ref{fig:enter-label-opt} illustrates the intensities versus photon excitation energy derived from electronic structures specifically in the context of incident light polarized along selected crystallographic planes of g-C$_3$N$_4$. The analysis reveals notable anisotropy in light absorption regardless of the chosen exchange-correlation functional. In the case of g-C$_3$N$_4$, the absorption spectrum when light is polarized along the 110 crystalline plane exhibits maximum absorption peaks between 450-500 nm. 
The outcomes obtained through DFTB analysis illustrate the electron densities associated with orbitals engaged in light absorption. This representation offers valuable insights into the subsequent electronic redistribution that occurs post-light absorption. Additionally, notable high-intensity peaks within the 550-650 nm wavelength of optical absorption edge range serve as empirical evidence supporting the assertion that g-C$_3$N$_4$ exhibits sensitivity to the visible light spectrum.

\begin{figure}[t!]
    \centering
    \includegraphics[width=0.48\textwidth]{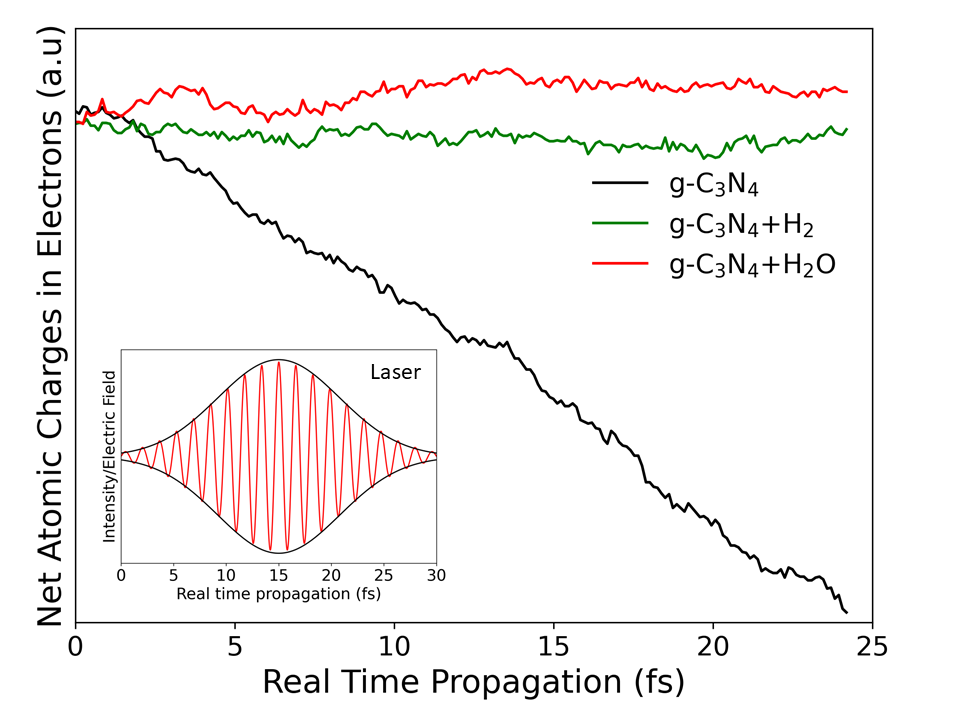}
    \caption{Net atomic charges as a function of laser pulse duration for the pristine case and with adsorbed molecules. 
    Laser pulse characterization is shown in the inset figure.
    These results illustrate differences in electronic behavior between the pristine surface and the system with adsorbed species by the applied laser pulse. }
    \label{fig:net-electron}
\end{figure}

In DFTB+ calculations, net atomic charges are determined through the 
Self-Consistent Charge (SCC) formalism, which iteratively adjusts the 
charge distribution to reflect interactions between atoms in a bonded 
system. These charges are calculated using Mulliken population 
analysis, where electron density is partitioned among atoms based on 
their molecular orbitals. In optical adsorbance calculations, 
the net atomic charge represents the difference between the number
of electrons an atom would possess in its neutral state and the 
redistributed electron density after bonding. 
Therefore, in Fig \ref{fig:net-electron} we present results 
for the net atomic charges as a function of laser pulse
duration for the pristine case and with the adsorbed molecules, that 
could reveal how the charge distribution evolves under laser 
irradiation. As the laser pulse
interacts with the material, it may induce electronic excitations, 
charge redistributions, and possibly ionization, depending on the 
material and laser characteristics. 
For the pristine case, we noticed that our results follow the experimental shape of the optical adsorbance reported in the literature \cite{Zhang2024,TONG2023116805}. 
Here, a water molecule 
adsorbed by the material keeps the net atomic charge constant 
to the presence of oxygen atom and the C-H bonding.

\subsection{Electron transport and conductance}

\begin{figure}[b!]
    \centering
    \includegraphics[width=0.48\textwidth]{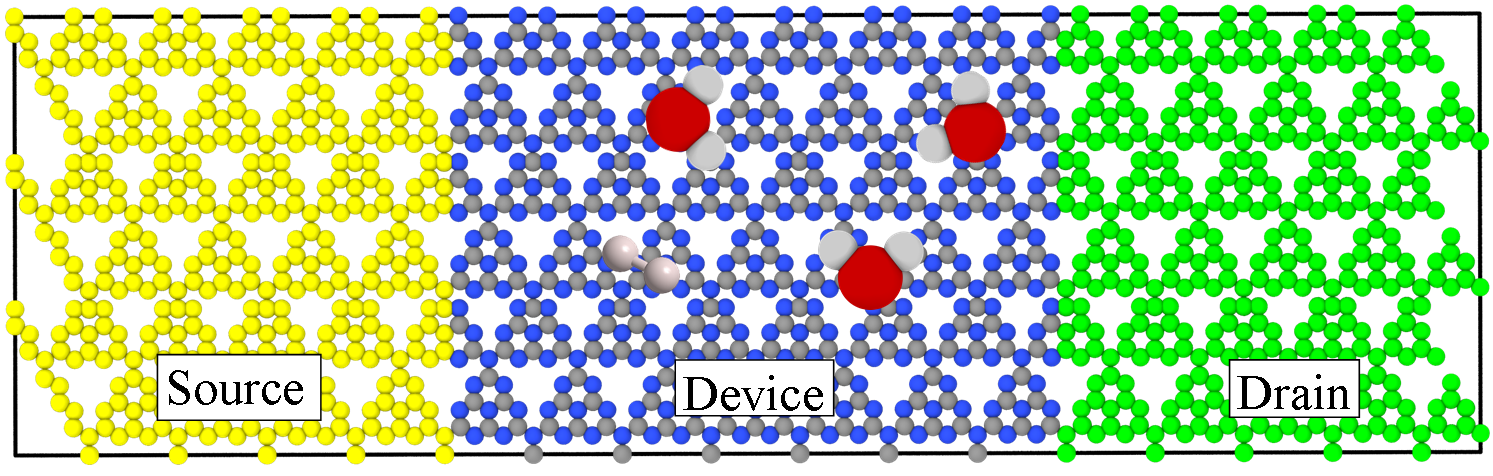}
    \caption{Geometric Configuration of g-C$_3$N$_4$ ribbon with source, device and drain regions involved in the electron transport calculations. To simulate realistic conditions and assess the impact of specific molecules on electron transport, H$_2$ and H$_2$O molecules were introduced into the device region.}
    \label{fig:transport}
\end{figure}

The electron transport calculations are conducted 
using the Non--Equilibrium Green's Functions (NEGF) 
formalism, a robust theoretical framework implemented 
in the DFTB code \cite{Pecchia_2008}. In Fig. 
\ref{fig:transport}, we present the schematics
the geometric configuration of the g--
C$_3$N$_4$ ribbon structure for electron transport calculations. 
To ensure accuracy and reliability, we follow several 
steps: 1) The structures are meticulously divided into 
distinct sections, including the principal layers, two 
electrode contacts (drain and source), and the device 
region. This partitioning facilitates a systematic
analysis of electron transport within the designated'scattering region'; 2) The drain section represents
the region where electrons exit the device, while the
source section corresponds to the region where
electrons enter; 3) A bias voltage of 1.0 V is applied
between the source and drain regions to investigate the current--voltage characteristics \cite{aligayev2024computational};
4) To simulate realistic conditions and investigate the impact of
specific molecules on electron transport, H$_2$ or 
H$_2$O molecules are introduced into the device
region. This allows us to study the interaction
between adsorbates and carbon-based materials,
observing their influence on the electron transport
properties.

\begin{figure}[b!]
    \centering
    \includegraphics[width=0.68\textwidth]{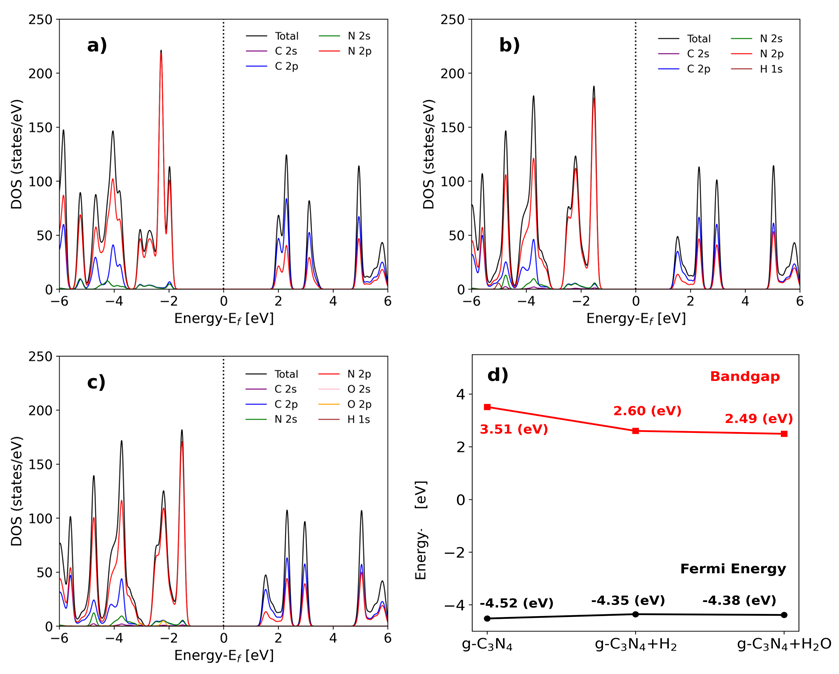}
    \caption{(Color online) The density of states for a) g-C$_3$N$_4$, b) g-C$_3$N$_4$+H$_2$, and c) g-C$_3$N$_4$+H$_2$O ribbon. The values of Bandgap and Fermi energy for each cases are shown in d).}
    \label{fig:DOS_ribbon}
\end{figure}
The density of states analysis 
for g-C$_3$N$_4$ and its interaction 
with H$_2$ and H$_2$O molecules, shown in Fig. \ref{fig:DOS_ribbon}, 
reveals notable effects on the electronic structure. For pristine 
g--C$_3$N$_4$, the DOS (Figure \ref{fig:DOS_ribbon}a) 
demonstrates a clear band gap of 3.5 eV  ( while a fully periodic system has a band gap of 2.67 eV), with dominant contributions from C--$2p$ and N--$2p$
orbitals near the Fermi level, confirming its semiconducting 
nature. Upon adsorption of H$_2$ (Fig. \ref{fig:DOS_ribbon}b),
the DOS remains largely similar, with minimal contribution
from the H--$1s$ orbitals, indicating weak 
interaction and a band gap 
decrease to 2.6 eV ( where a band gap of 2.71 eV was obtained for 
the fully periodic case). In contrast, the 
adsorption of H$_2$O (Fig. \ref{fig:DOS_ribbon}c) 
introduces a more significant change in 
the DOS, particularly due to the O--$2p$ 
orbitals contributing states below the 
Fermi level. This results in a further 
decrease in the band gap to 2.5 eV
(2.8 eV for the fully periodic system) and 
a notable shift in the Fermi energy. 
These findings, summarized in Fig. 
\ref{fig:DOS_ribbon}d), suggest that while both H$_2$ and H$_2$O 
adsorption affect the electronic 
properties of g--C$_3$N$_4$, the interaction with H$_2$O is stronger, 
leading to a more pronounced 
modification of the band 
structure. Nonetheless, the 
semiconducting nature of g-C$_3$N$_4$ is preserved in all cases.

\begin{figure}[b!]
    \centering
    \includegraphics[width=0.48\textwidth]{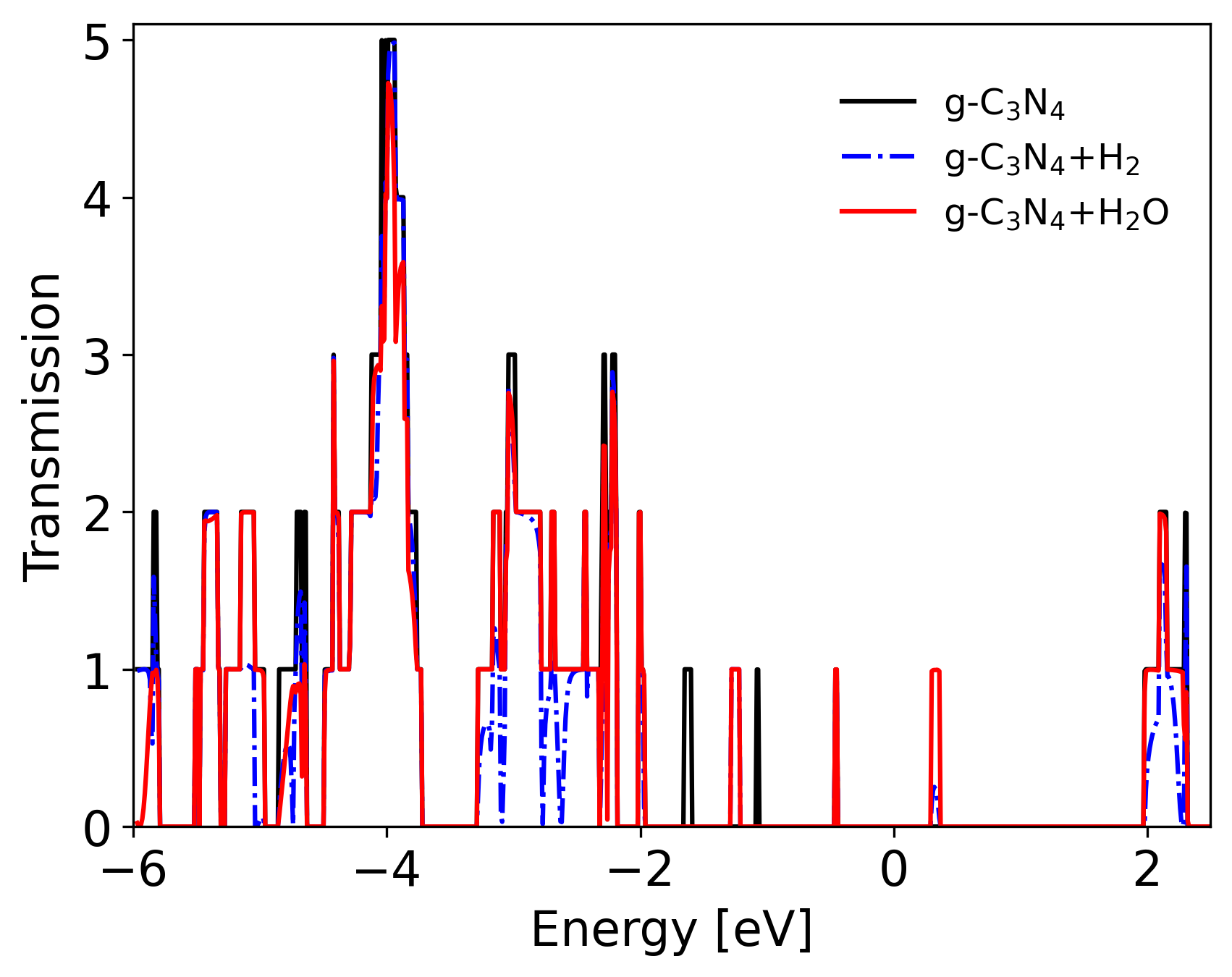}
    \caption{ Comparison of electron transmission probability for g-C$_3$N$_4$ with H$_2$ and H$_2$O molecules in different orientations.}
    \label{fig:transport2}
\end{figure}

The electron transmission for g-C$_3$N$_4$ and its interaction 
with H$_2$ and H$_2$O molecules, as depicted in Fig 
\ref{fig:transport2}, reveals key 
insights into the system's electronic 
transport properties. The pristine g--
C$_3$N$_4$ exhibits distinct 
transmission peaks, with gaps 
corresponding to its intrinsic density of states (See Fig 
\ref{fig:DOS_ribbon}a), particularly in negative 
region, where electron transmission is 
suppressed due to the absence of 
available electronic states within the 
material’s valence band. This 
suppression is a result of the 
material's band gap, preventing 
electrons from conducting in these 
energy ranges. Adsorption of H$_2$ and H$_2$O introduces 
minor shifts in the transmission peaks, 
suggesting localized electronic 
interactions between the molecules and 
the surface. However, the overall 
transmission characteristics remain 
largely consistent across different 
molecular orientations, indicating that 
the adsorbates have only a subtle 
influence on the electron transport. 
This suggests that while the molecular 
adsorption alters local electronic 
states, the fundamental transmission 
properties of g-C$_3$N$_4$ remain 
preserved.

\begin{figure}[t!]
    \centering
    \includegraphics[width=0.48\textwidth]{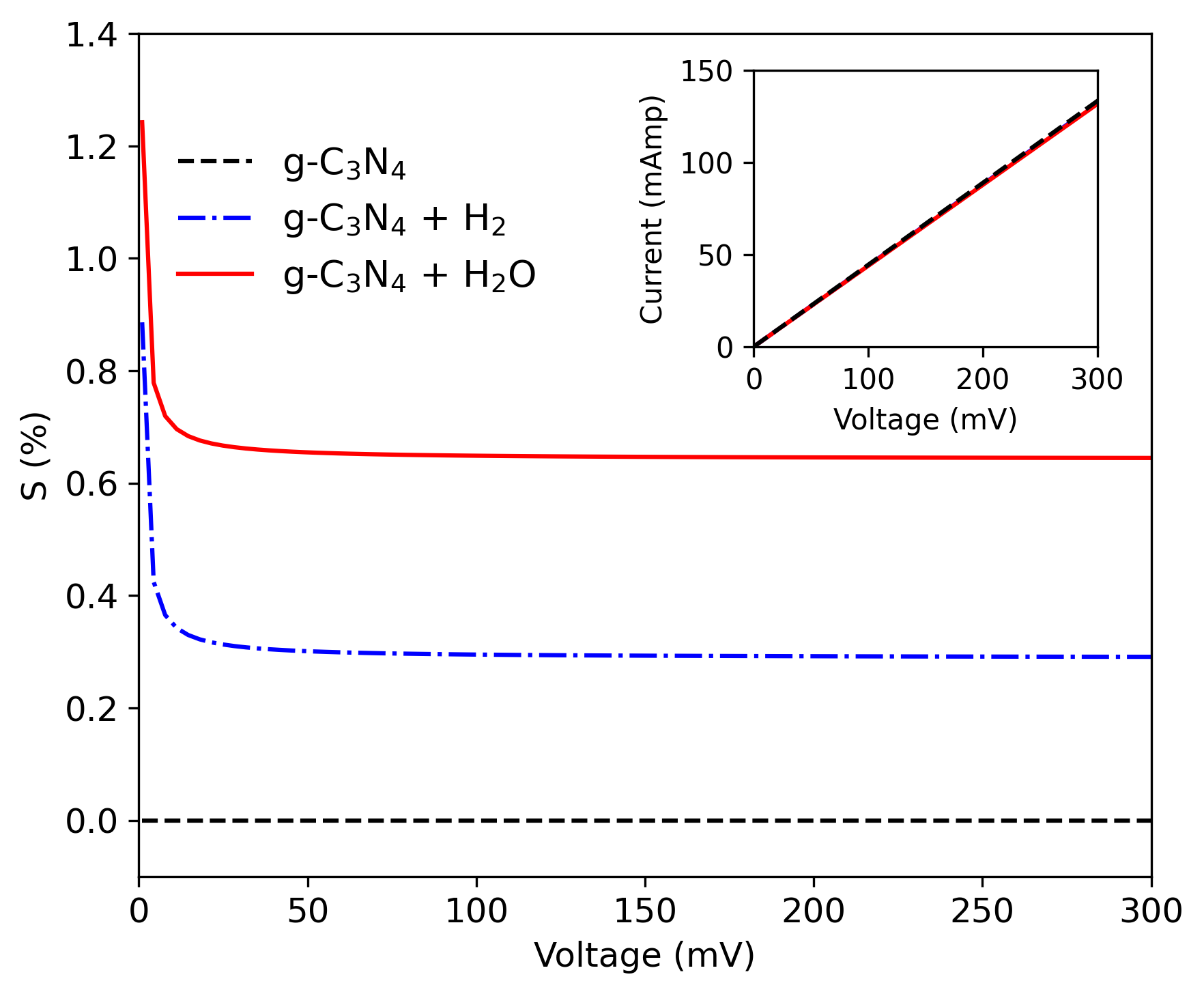}
    \caption{Sensitivity of the g-C$_3$N$_4$ sheet compared to adsorbed hydrogen and water molecules as a function of applied voltage. The inset shows the typical current–voltage (I–V) plot for each system, highlighting that the presence of a water molecule decreases the current in the g-C$_3$N$_4$ sheet.}
    \label{fig:U-V_plot}
\end{figure}

Fig \ref{fig:U-V_plot} shows the difference 
of the current for each molecule $X$ as a function of
the voltage as:
\begin{equation}
    S = 100 \% \frac{\big| I_{X}-I_g \big|}{I_{X}},
\end{equation}
with $I_{X}$ of each molecule and the surface and 
$I_g$ the current of the pristine and Ni-doped graphene 
\cite{aligayev2024computational}. 
The inset plots display the tunneling currents of the sheets as
a function of voltage for different adsorbed molecules,
represented by I–U characteristic graphs for voltages below
300 mV.
Our results show a sensitivity increase of 0.38\% for hydrogen 
molecules and 0.70\% for water upon molecular doping, with water 
exhibiting the highest sensitivity and a significant effect on the 
current–voltage response. This enhanced sensitivity is attributed to 
charge polarization induced by oxygen, which modifies the electronic 
properties of the sheet.

\subsection{
Material's Parameters for Photocatalytic Efficiency}

High visible light absorption is an essential property for efficient photocatalytic applications. To address this, we conducted additional calculations for the unit cell and the $2 \times 2 \times 1$ supercell of g-C$_3$N$_4$, as well as its interaction with H$_2$O molecules. Specifically, for the unit cell, the HOMO energy was calculated as $-1.4995$ eV, the LUMO energy as $-0.2753$ eV, resulting in a HOMO-LUMO gap ($E_g$) of $1.2242$ eV. The chemical potential ($\mu$) was $0.8874$ eV, chemical hardness ($\eta$) was $0.6121$ eV, electronegativity ($\chi$) was $-0.8874$ eV, and electrophilicity ($\omega$) was $0.6433$.  
For the $2 \times 2 \times 1$ supercell of g-C$_3$N$_4$, the HOMO energy was $-3.6985$ eV, the LUMO energy was $-2.5055$ eV, with an $E_g$ of $1.193$ eV. The corresponding $\mu$, $\eta$, $\chi$, and $\omega$ were $3.1020$ eV, $0.5965$ eV, $-3.1020$ eV, and $8.066$, respectively.  
Additionally, for the $2 \times 2 \times 1$ supercell with an adsorbed H$_2$O molecule, the HOMO energy was $-3.6691$ eV, the LUMO energy was $-2.4803$ eV, with an $E_g$ of $1.1888$ eV. The values of $\mu$, $\eta$, $\chi$, and $\omega$ were calculated as $3.0747$ eV, $0.5944$ eV, $-3.0747$ eV, and $7.9524$, respectively.  
These results provide 
chemical properties of g-C$_3$N$_4$ and highlight its potential as an effective photocatalyst.

\begin{figure*}[t!]
    \centering
    \includegraphics[width=0.95\textwidth]{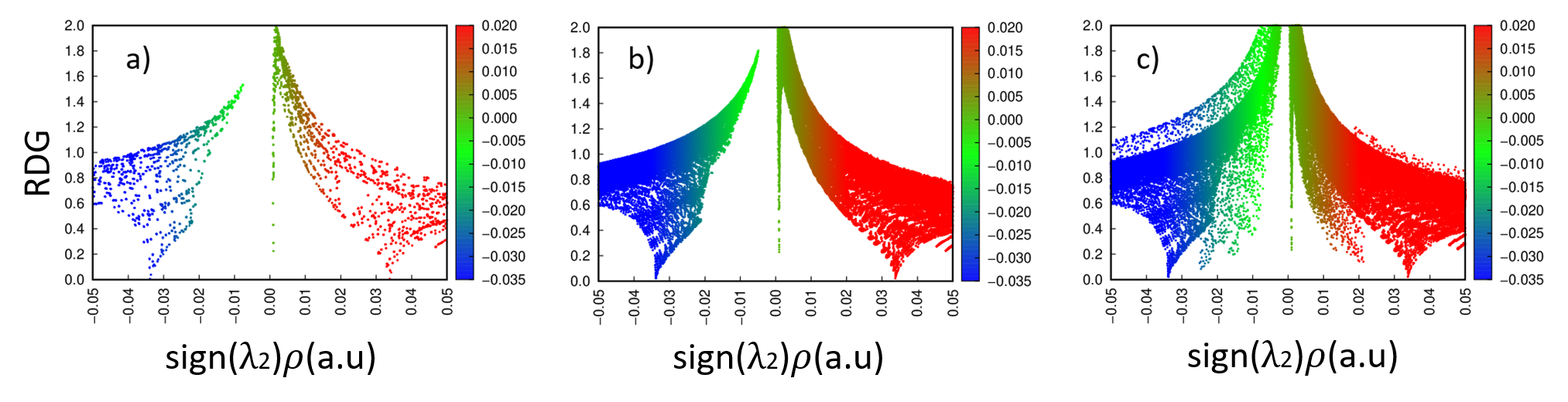}
    \caption{ a) Reduced Density Gradient of the unit cell of g-C$_3$N$_4$, b) the supercell of g-C3N4, and c) the H2O molecule on the surface supercell of g-C$_3$N$_4$.}
    \label{fig:rdg}
\end{figure*}

Fig.~\ref{fig:rdg} presents the Reduced Density Gradient (RDG) 
analysis as a function of $\mathrm{sign}(\lambda_2)\rho$ \cite{ALIGAYEV2024162022} (in 
atomic units) for three systems: (a) the unit cell of g-
C$_3$N$_4$, (b) the supercell of g-C$_3$N$_4$, and (c) the 
H$_2$O molecule adsorbed on the surface of the g-C$_3$N$_4$ 
supercell. The vertical axis represents the RDG values, while 
the horizontal axis corresponds to $\mathrm{sign}
(\lambda_2)\rho$, with a color gradient indicating the 
electron density ($\rho$) values, ranging from blue (low 
density) to red (high density).  
Fig. \ref{fig:rdg}-a) illustrates the RDG distribution for the 
unit cell of g-C$_3$N$_4$, with points predominantly located 
in the blue and green regions. This distribution signifies 
weak non--covalent interactions, such as van der Waals forces, 
with minimal evidence of strong attractive interactions. In 
Fig. \ref{fig:rdg}-b), the RDG profile for the g-C$_3$N$_4$ 
supercell exhibits a broader distribution, extending into the 
green and red regions. This indicates an increase in both weak 
non-covalent interactions and stronger attractive forces, 
attributed to the larger system size and enhanced spatial 
interaction among atoms.  
In Fig. \ref{fig:rdg}-c), the RDG analysis of the H$_2$O 
molecule adsorbed on the g-C$_3$N$_4$ surface reveals new 
interaction features. The extended spread of points in the 
blue and green regions highlights weak hydrogen bonding and 
charge redistribution effects induced by adsorption. 
Additionally, the presence of data points in the red region 
suggests localized strong attractive forces between the H$_2$O 
molecule and the g-C$_3$N$_4$ surface.

The exciton binding energy ($E_b$), exciton radius ($R_e$), 
and exciton energy of g-C$_3$N$_4$ can be calculated using the 
Mott-Wannier model~\cite{D0TC02066D,Shahrokhi,D2CP03410G}, 
which depends on the relative dielectric permittivity 
($\varepsilon_r$) and the reduced mass of the exciton ($m^*$). 
The exciton energy, which corresponds to the energy difference 
between the conduction band minimum and the valence band 
maximum minus the binding energy of the exciton, is expressed 
as: E$_{\text{exciton}}$ = E$_{\rm g}$ - E$_{\rm b}$
where $E_g$ is the electronic bandgap of the material, and 
$E_b$ is the binding energy of the exciton. The exciton 
binding energy and radius are defined as: 
E$_{\rm b}$ = E$_{\rm H}$ $(m^*_r/\varepsilon^2_r)$ and R$_e$ = $a_0$$(\varepsilon_r/4m^*_r)$, 
where $a_0$ is the Bohr radius, $E_H$ is the Rydberg energy 
(13.6~eV), and $m^*_r$ is the reduced mass of the exciton, given 
by:$\frac{1}{m^*_r} = \frac{1}{m_e^*} + \frac{1}{m_h^*}$,
with $m_e^*$ and $m_h^*$ denoting the effective masses of the 
electron and hole, respectively. For  g-C$_3$N$_4$, 
determining these parameters effective electron and hole 
masses and the dielectric constant enables an accurate 
estimation of the exciton binding energy, which is essential 
for understanding its optical and electronic properties. 
A low exciton binding energy would indicate efficient charge separation, making  g-C$_3$N$_4$ a promising material for 
photocatalytic and optoelectronic applications with  R$_{\rm e}$ = 2.88 $\AA{}$, E$_{\rm b}$ = 367 eV and $m_e^*$, 
$m_h^*$ ($m_e$) equals 0.4, and 1.4 respectively.

\section{Dissociative mechanisms for water splitting}

By utilizing a reactive molecular beam with high kinetic energy, gas-surface reactions involving high activation barriers can be accelerated. In a molecular beam setup, gas molecules are directed in a controlled stream towards a target surface, providing precise control over kinetic energy, angle, and molecular orientation. This method enables overcoming reaction barriers that would typically require elevated pressures or temperatures when using a random gas in its natural state. In the case of water in its gas phase, the molecular beam allows for more effective interaction with the surface by enhancing the probability of dissociation or reaction, thus facilitating processes like water splitting or catalysis.
Thus, to investigate the adsorption dynamics and potential 
dissociation of H$_2$O molecules interacting with the g--C$_3$N$_4$ sheet, we performed 
molecular dynamics (MD) simulations using a $2\times2\times1$ 
supercell. The simulations utilized both the SCC-DFTB method and the 
ReaxFF force field. Prior to adsorption studies, the surface was 
optimized and equilibrated at 300 K using a Nosé-Hoover thermostat. 

For 
the adsorption dynamics, a target area of 1 nm$^2$ was defined on the 
surface, and molecules were randomly distributed using the velocity 
Verlet algorithm. H$_2$O molecules were emitted vertically with random 
orientations from an initial distance of 0.8 nm above the surface, with 
an impact energy of 5 eV relative to the center of mass.
We generated 1,000 independent trajectories for each molecule, using a 
time step of 0.25 fs, with simulations running for 350 fs. This 
simulation time was chosen to ensure convergence while allowing the 
molecules to interact with the carbon sheets and remain bonded. The 
timeframe was carefully selected to balance the prevention of 
detachment and achieve convergence in our MD simulations.
This approach has been previously employed in studies of hydrogenation mechanisms of fullerene cages \cite{DOMINGUEZGUTIERREZ2018189}, the electronic properties of borophene \cite{C7TC00976C}, and dynamic physisorption pathways of molecules on alumina surfaces \cite{aluminadftb}, yielding results in excellent agreement with first-principles DFT calculations.

\begin{figure}[t!]
    \centering
    \includegraphics[width=0.78\textwidth]{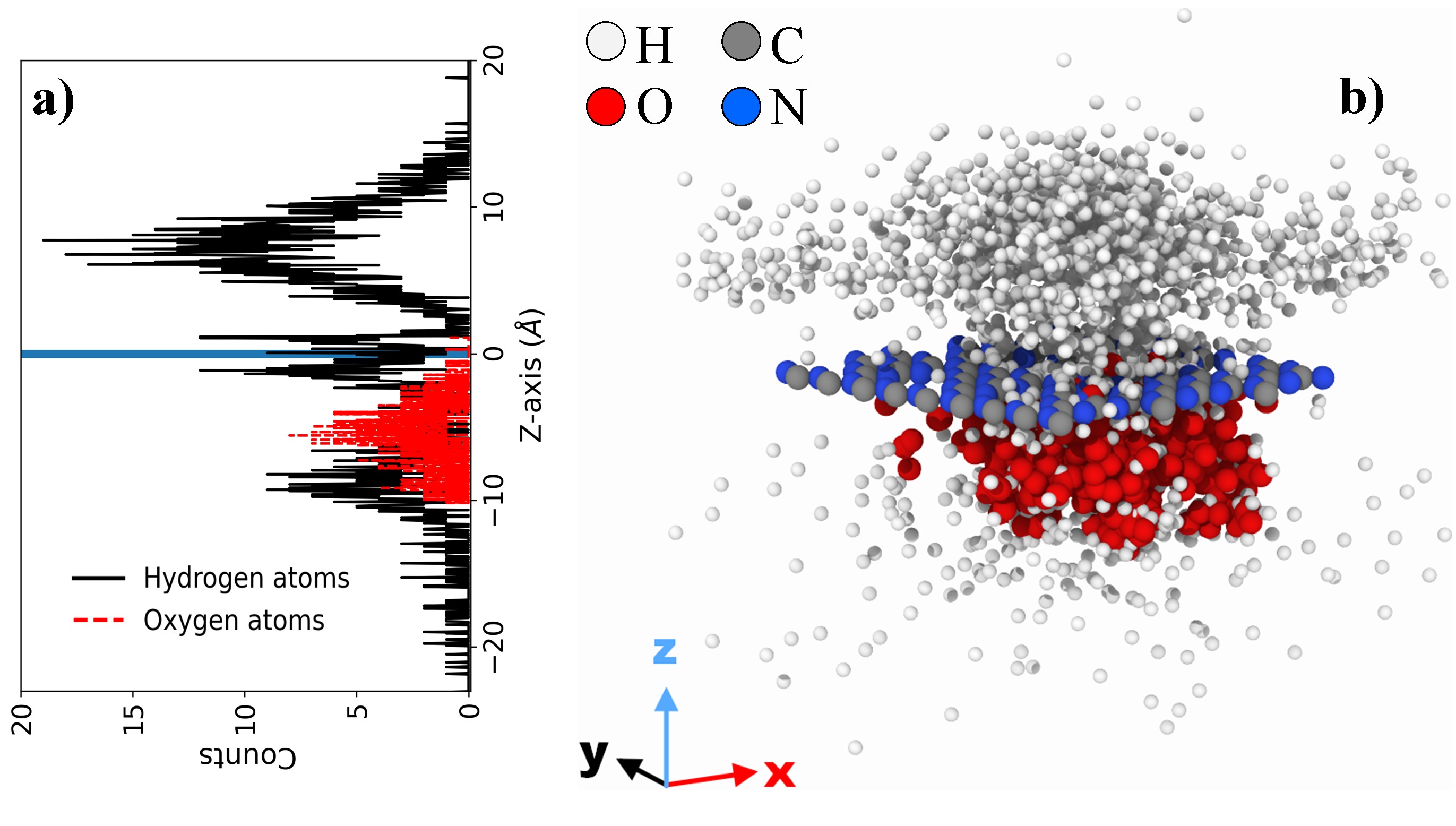}
    \caption{a) Histogram shows the position of H and O atoms from the surface of g-C$_3$N$_4$ and b) Visualization of H$_2$O splitting at the end of dynamics. }
    \label{fig:dynamics}
\end{figure}

In Fig. \ref{fig:dynamics}, we present the results from scc-DFTB 
simulations, showing the final frame of the molecular dynamics to 
illustrate the production of hydrogen atoms following the collision of 
water molecules with the g-C$_3$N$_4$ sheet. The dissociation pathways 
observed are as follows: i) H$_2$+O for 6.5\% of the cases, ii) OH+H 
for 12.7\%, iii) O+H+H for 78\%, and iv) no dissociation for 2.8\% of 
the events. These results are in good agreement with ReaxFF 
simulations, indicating that the atomic configuration and geometry of 
the sheet primarily promote water dissociation into O+H+H, followed by 
OH+H. This demonstrates the material's potential for water-splitting 
applications and hydrogen production.
These observations are directly influenced by the impact energy of the water molecules, set at 5 eV. If this energy is decreased, we observed that the primary dissociation pathway shifts, with the water molecule predominantly splitting into HO+H. 
In addition, the results obtained from reaxFF simulations are 
consistent with those reported in the supplementary material. This 
consistency opens up the opportunity to expand the simulation scale, 
increasing the number of atoms and water molecules to further 
investigate the interaction of this material with water in gas and 
liquid phases, which could significantly affect its electronic 
properties.

\section{Concluding remarks}
\label{sec:concl}
We employ computer simulations to explore the adsorption and 
separation mechanisms of H$_2$O molecules and their impact on 
the optical and electronic properties of g-C$_3$N$_4$ sheets. 
Our approach utilizes a multi-scale methodology, incorporating 
DFT, SCC-DFTB, and ReaxFF methods, complemented by van der 
Waals corrections that are essential for 
accurately capturing the dynamics of dissociation, 
chemisorption, and molecular formation.
Our findings demonstrate that g-C$_3$N$_4$ exhibits 
significant 
advantages in the separation of H$_2$O molecules compared to a 
variety of other 2D nanomaterials. Notably, it effectively 
dissociates H$_2$O into H$_2$ and O$_2$, achieving the highest 
probabilities for separation. The inherent porosity of g--
C$_3$N$_4$ not only enhances gas separation rates but also 
facilitates hydrogen production from H$_2$O as O+H+H. 
Furthermore, electron transport calculations conducted using 
the non-equilibrium Green's function method indicate that the 
presence of H$_2$O leads to a slightly decrease in the 
conductance of g-C$_3$N$_4$ materials. 
Overall, the RDG analysis underscores the 
dominance of weak non-covalent interactions in pristine g-
C$_3$N$_4$ systems. However, the adsorption of H$_2$O enhances 
both weak and strong interactions, providing insights into the 
reactivity and interaction mechanisms at g-C$_3$N$_4$ 
surfaces.

Moreover,improving the adsorption ability of 
g-C$_3$N$_4$ can be achieved through several strategic 
approaches that can be modeled by our multiscale approach. 
Engineering porosity by optimizing pore size and distribution 
increases 
the accessibility of adsorption sites, while defect 
engineering introduces structural irregularities, such as 
vacancies or edge modifications, that act as highly reactive 
sites for stronger H$_2$O binding. External influences, like 
applying an electric field or external bias, can polarize 
H$_2$O molecules and the surface, strengthening adsorption 
interactions and lowering dissociation barriers. Furthermore, 
optimizing operating conditions, such as temperature and 
pressure, can enhance adsorption kinetics and interaction 
dynamics.  Additionally, combining g-C$_3$N$_4$ with other 
materials, such as metal oxides, graphene, or transition--
metal dichalcogenides, can create hybrid structures with 
enhanced 
charge transfer and improved catalytic properties.
These combined strategies provide a comprehensive pathway for 
significantly improving the adsorption capabilities of g--
C$_3$N$_4$ for water splitting and other catalytic 
applications that will be modeled in our future work.
These insights underline the potential of g-C$_3$N$_4$ as a promising material for applications in gas separation and hydrogen production, as well as the necessity for further investigation into its electronic properties under varying environmental conditions.


\section*{Acknowledgements}
We acknowledge support from the European Union Horizon 2020 research
 and innovation program under grant agreement no. 857470 and from the 
 European Regional Development Fund via the Foundation for Polish 
 Science International Research Agenda PLUS program grant 
 No. MAB PLUS/ 2018/8. We gratefully acknowledge Polish high-performance computing infrastructure PLGrid (HPC Center: ACK Cyfronet AGH) for providing computer facilities and support within computational Grant No. PLG/2024/ 017084

 \bibliographystyle{elsarticle-num} 
 \bibliography{biblio}





\end{document}